\documentclass[twocolumn,aps,amsmath,showpacs,pre]{revtex4}
\usepackage{graphicx}
\usepackage{amssymb}
\usepackage{amsmath}
\usepackage{color}
\DeclareMathOperator{\sech}{sech}

\begin{document}

\title{Using Nanoresonators with Robust Chaos as HRNGs}

\author{Lucas~R.~Rodrigues,$^{2}$ W.~G.~Dantas,$^{1}$ Sebastian~Ujevic,$^{1}$ and  A.~Gusso,$^{1}$} 

\affiliation
{$^{1}$Departamento de Ci\^encias Exatas, UFF, EEIMVR, Brazil \\
$^{2}$Instituto de F\'isica, UFF, Brazil}

\date{\today}

\begin{abstract}
In this paper, we investigate theoretically the potential of a nanoelectromechanical suspended beam resonator excited by two-external frequencies as a hardware random number generator (HRNG). This system exhibits robust chaos, which is usually required for practical applications of chaos. Taking advantage of the robust chaotic oscillations we consider the beam position as a possible random variable and perform tests to check its randomness. The beam position collected at fixed time intervals is used to create a set of values that is a candidate for a random sequence of numbers.  To determine how close to a random sequence this set is we perform several known statistical tests of randomness. The performance of the random sequence in the simulation of two relevant physical problems, the random walk and the Ising model, is also investigated. An excellent overall performance of the system as a random number generator is obtained. 
\end{abstract}


\maketitle

\section{Introduction}
\label{intro}

Random numbers are required in many practical applications such as cryptography  for secure data storage or  communications~\cite{nien07,behnia08} and simulations of stochastic processes~\cite{binder}. Some applications may rely upon the use of pseudo-random numbers, generated by deterministic algorithms, while others, may require real random numbers generated by hardware random number generators (HRNGs) (or physical random number generators) based on fundamental physical processes. 

Independently of its source of generation, a sequence of values is defined as random if the numbers composing it have the same statistical properties observed in an infinite random sequence~\cite{knuth69}. Among such properties, we can highlight, the homogeneity in the distribution of values and its mutual independence, i.e. each number in the sequence has a value completely uncorrelated with those already present and with those that still will be included~\cite{mariotania}. Although the task of defining what is a random sequence is quite simple, to determine if a given set of numbers can be classified as truly random is not. In fact, this is an unsolvable problem, because there is no finite set of tests capable of determining if a sequence of values is genuinely random. Instead, tests applied in a sequence can only disqualify it as a truly random set of values.

To artificially generate random numbers we can use deterministic algorithms that can produce sequences with the desirable statistical properties. Those algorithms are examples of pseudo-random number generators (PRNGs). However, no matter how good the algorithm can be, all of them display a failure: after a certain number of values generated the sequence repeats itself. The amount of non-repeated numbers is the period of the generator and in a good PRNG it must be as long as possible. However, sometimes even a generator displaying very long periods and passing by several statistical tests could still fail when used in some other application. One example of such a case was reported by Ferrenberg et. al.~\cite{ferrenberg92}, who demonstrated that a well-known and tested PRNG still kept enough correlation among the values produced and failed when used to simulate the Ising model.

In practice, for applications in HRNGs, some phenomena producing real random variables are extremely hard to control and therefore to be used to generate a random sequence. Some of these phenomena are the radioactive decay process~\cite{alkassar05}, time arrival of particles in the atmosphere brought by cosmic rays~\cite{wu17} and thermal noise~\cite{lavine17}. However, alternative physical sources of randomness can be used. For instance, HRNGs have been implemented using jitter noise of clock signals and metastability in circuits~\cite{yang14,kim17,kuan14} or thermal or shot noises obtained from analog electronic circuits~\cite{petrie00,figliolia16}. 

Since the work of Ulam and von Neumann~\cite{ulam47}, a new class of methods, based upon systems displaying a chaotic dynamics,  was tested as PRNGs and, more recently, implemented in HRNGs. Because such systems have a strong dependence on the initial condition and their evolution is usually quite erratic they could be  excellent candidates to emulate a random process. Also, because a chaotic motion is, by definition, non-periodical, we should not have any concern about the repetition of values. Simple iteration equations (maps) are among the dynamical systems that can exhibit chaos. Their usefulness as a PRNG was positively confirmed, for instance, for the logistic map~\cite{feigenbaum78} by Phatak and Suresh~\cite{phatak}. Similar results were also observed for other kinds of maps~\cite{gonzalez02}. HRNGs based upon different maps have been implemented in the form of electronic circuits.

Continuous time chaotic signals (flows) can also be used as a physical source of randomness. Electronic circuits based upon Chen's systems, LC based chaotic oscillators and jerk circuits have been implemented and tested successfully~\cite{wan18}. Another class of relevant physical systems that can display chaos are micro/nano-electromechanical (MEMS/NEMS) resonators~\cite{amorim15,dantas18,gusso3,wang98,karabalin09,barcelo19}. MEMS/NEMS are generally considered as electromechanical alternatives to purely electronic circuits. Their advantages over electronic devices usually are their small size and low power consumption. Due to their smallness they can also achieve very high frequencies of oscillation~\cite{younis11}. For many applications, like in mobile communications, these are very important features. For this reason MEMS/NEMS resonators in many different configurations have been investigated as sources of chaotic signals~\cite{wang98,karabalin09,barcelo19}.   

As it is the case for most of known continuous chaotic systems, chaos in the investigated MEMS/NEMS resonators is expected to be fragile. That means, any changes in the system parameters may cease the oscillations in the chaotic regime~\cite{zeraoulia12}. In the relevant parameter space, regions of chaos are intermingled with those of periodic behavior or other attractors, such as the escape to infinity. It was verified in Refs.~\cite{amorim15,gusso3,gusso19c} that chaos is fragile in suspended beam MEMS/NEMS resonators in the most usual configurations and operational conditions.  However, for practical applications, robust chaos is generally required~\cite{zeraoulia12,kocarev01}. Robust chaos is defined by the persistence of the chaotic attractor as the parameters of the system vary~\cite{zeraoulia12}. Fortunately, Gusso et. al.~\cite{gusso3} have demonstrated that a doubly clamped suspended beam resonators with two lateral electrodes exhibit robust chaos when actuated by two AC voltages with distinct frequencies. The system thus becomes a strong candidate as a source of randomness.

Our goal in this paper is to investigate the potential of this particular NEMS resonator as a HRNG. We organized our manuscript as follows. In Section~\ref{model}, we will define the system and the model for its dynamics, as well as the approximation involved to obtain a proper equation of motion describing it. Section~\ref{prng} is used to discuss how we collect a series of values and compose a sequence that will have its randomness evaluated. In Section~\ref{secrand}, we apply a series of tests divided into two categories: the statistical ones and the physical ones, where we used a set of numbers obtained for a particular point of the space parameter for which the dynamic is chaotic. Section~\ref{other} will be devoted to evaluate what happens with other points of the parameter space where the set of values obtained has not a good performance. Also, in this section, we propose a method to improve the generation of random numbers through a shuffle protocol. Finally, in Section~\ref{conc} the conclusions are presented as well as possible extensions to our work.

\section{System Model}
\label{model}

\begin{figure}[t!]
\begin{center}
\includegraphics[scale=0.35]{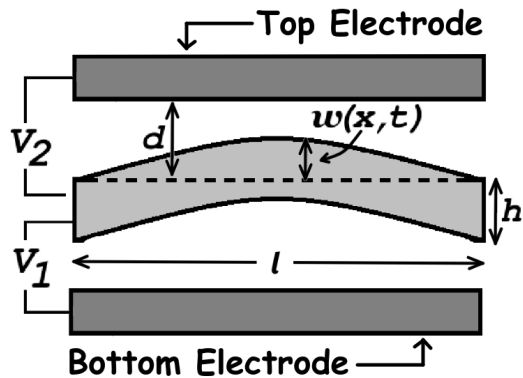}
\caption{Schematic diagram showing a lateral view of the beam resonator with two-sided electrodes..}
\label{fig1}
\end{center}
\end{figure}

In this section we briefly review the physical and mathematical model of the system. More details can be found in Refs.~\cite{dantas18,gusso3}. For the purpose of the numerical simulations, we are going to consider a realistic NEMS resonator. We consider the beam to have a constant rectangular cross-section of thickness $h$ and width $b$ along its length $l$, Fig.~\ref{fig1}. We also consider a slender beam (large $l/h$ ratio) with homogeneous and isotropic elastic properties (a constant Young modulus). Regardless of the boundary conditions, such a system could be modeled by the Euler-Bernoulli beam theory~\cite{graff75}. 

However, because in the chaotic regime a beam clamped at both ends can be subject to large transverse displacements, compared to its thickness, it is mandatory to include the effect of the mid-plane stretching~\cite{younis11}. It is responsible for a nonlinear hardening effect on the elastic restoring force. The electrostatic force due to the applied voltages $V_1(t)$ and $V_2(t)$ through the electrode gaps is modeled considering that the beam suffers a small bending, Fig.~\ref{fig1}. This is justified whenever the beam vibrates in its lowest order modes and with amplitudes that are small compared to its length. This is the case in our system since the amplitudes are limited by the gaps of dimension $d$ which are going to satisfy $d \ll l$. In this case, the beam can be assumed as piece-wise plane and the electrostatic force can be approximated as that between two parallel plates along each infinitesimal segment of the beam. Finally, we also assume, as usually done, that dissipation occurs due to a viscous damping, proportional to the local velocity of the beam. Nonlinear damping is generally expected in such systems~\cite{zaitsev12,gusso16,gusso19b} however, so far, most of the models of dissipation can only be applied to systems vibrating periodically and we ignore them here.

Considering all the above assumptions, the partial differential equation modeling the system results to be~\cite{amorim15,dantas18}

\begin{eqnarray}
\label{PDE}
EIw'''' + \rho A \ddot{w} + c\dot{w} - \left( \frac{EA}{2l} \int_0^l {w'}^2 dx \right) w'' \nonumber \\
+ \frac{\epsilon_0 b}{2} \left[ \frac{ V_1(t)^2 }{(d+w)^2} - \frac{ V_2(t)^2 }{(d-w)^2 } \right] = 0 \,.
\end{eqnarray}

\noindent In this equation $w(x,t)$ corresponds to the vertical displacement along the beam, comprised between $x=0$ and $x=l$, and subject to the boundary conditions $w(0,t)=w(l,t)=w'(0,t)=w'(l,t)=0$.  The over-dots and primes represent derivatives with respect to time ($t$), and space ($x$), respectively.  $E$ denotes the Young modulus, $I=b h^3/12$ the geometric moment of inertia, $\rho$ the beam density, $A=b h$ its cross-sectional area, $c$ corresponds to the linear damping coefficient, $\epsilon_0 = 8.85 \times 10^{-12}$~F/m corresponds to the vacuum permittivity. In Eq.~(\ref{PDE}), the first two terms correspond to the elastic and inertia terms of the Euler-Bernoulli beam theory and the third term to the viscous damping. The term proportional to $w''$ corresponds to the nonlinear restoring force due to the mid-plane stretching. The last term gives the contribution of the electrostatic force.

We do not solve Eq.~(\ref{PDE}) directly. Instead, we work with a reduced order model with a single degree of freedom. It has already been shown, both theoretically and experimentally, that the nonlinear and chaotic dynamics of a beam or a string driven by frequencies close to that of a given mode can be very well described using a reduced order model that includes only this mode of vibration~\cite{moon79,moon83,bajaj92}. We consider only the first mode because actual devices usually operate at the first resonant mode since it provides the best performance for excitation and read-out of the oscillations~\cite{uranga15}.

The reduced order model is obtained, applying the Galerkin method~\cite{younis11} to Eq.~(\ref{PDE}). We approximate $w$ by $w(x,t) = u(t) \phi_1(x)$, where $\phi_1(x)$ denotes the base function which corresponds to the first mode-shape of a doubly clamped beam described mathematically by the Euler-Bernoulli equation, given by the first two terms in Eq.~(\ref{PDE}). Using the orthonormality of the mode-shapes and performing a suitable change of variables we obtain the following non-dimensional nonlinear ordinary differential equation (for more details on the derivation see Refs.~\cite{dantas18,amorim15})

\begin{equation}
\label{odeins}
\ddot{s} + \beta \dot{s} + s + \alpha s^3 + F^e(s,\tau)  = 0 \, .
\end{equation}

\noindent In this equation the  variable $s = s(\tau)$ can be understood as the approximate non-dimensional displacement of the beam at the position of maximal amplitude. It is related to $w$ by $s(\tau) = w_{max}(\tau)/d$ where $w_{max}(\tau)=w(x=0.5 l,\tau)$ corresponds to the maximum beam displacement that occurs at  the beam center. The time derivatives are now with respect to the non-dimensional time $\tau = t/\omega_0$, where $\omega_0$ denotes the natural frequency of the first mode. The cubic non-linearity term is simply $\alpha = 0.719 (d/h)^2$. The damping factor $\beta$ is related to the quality factor $Q$ simply by $\beta = Q^{-1}$. The last term corresponds to the electrostatic force, given by  

\begin{eqnarray}
\label{Fes}
F^e(s,\tau) &=& 1.218 \frac{\epsilon_0 b l}{2 k_{eff} d^3} \times \nonumber \\ 
& & \biggl[ V_1^2(\tau) \int_0^1 \frac{\phi_1(x')}{(1+\phi_1(x') s(\tau)/d)^2} dx' \nonumber \\
& & -V_2^2(\tau) \int_0^1 \frac{\phi_1(x')}{(1-\phi_1(x') s(\tau)/d)^2} dx' \biggr] \, \nonumber \\
&=& B \left[ V_1^2(\tau) I^e(s(\tau)) - \left( V_2(\tau) \right)^2 I^e (-s(\tau))  \right] \, ,
\end{eqnarray}

\noindent where $B = 0.609 \epsilon_0 b l/(k_{eff} d^3)$, with $k_{eff} = 384 E I/l^3$ denoting the effective elastic constant of the beam. 

To circumvent the time consuming numerical calculation of the integrals in $F^e$ what we have been doing~\cite{amorim15,dantas18}  to solve Eq.~(\ref{odeins}) numerically in an efficient manner is to replace $I^e$ by a suitable approximate function of the form 

\begin{equation}
\label{Approx}
I_a^e(s) =  \frac{a_0}{(1+\sum_{i=1}^3 a_i s^i)} \, .
\end{equation}

The coefficients $a_i$ assume the values $a_0 = 0.829700$, $a_1 = 1.521728$, $a_2 = 0.389925$, and $a_3 = -0.104225$, and result in an accuracy of  $0.6\%$ compared to the numerical evaluation of the integrals over the range   $-0.7 \leq s \leq 0.7$, which is the relevant range of $s$  in the numerical solution of Eq.~(\ref{odeins}).

In order to obtain robust chaotic dynamics, the resonators are actuated by DC and AC voltages. A constant voltage bias, $V_{DC}$, is applied between the beam and both lateral electrodes. It is the main responsible for changing the effective potential felt by the beam. For small $V_{DC}$, the system has a single-well potential. However, as it increases, a double-well can form.  The electrodes are also excited by alternate signals superposed to the DC voltage. In order to have robust chaos the alternate signals must have distinct frequencies~\cite{gusso3} and we take $V_1(\tau) = V_{DC} + V_{AC} \cos(\zeta_1 \tau)$ and $V_2(\tau) = V_{DC} + V_{AC} \cos(\zeta_2 \tau)$, where $\zeta_i = \omega_i/\omega_0$. We have considered that the AC voltages applied to the two electrodes have the same amplitude and phase, just in order to decrease the number of free parameters. However, the normalized frequencies of excitation can be different. In what follows we assume that the normalized frequencies $\zeta_1$ and  $\zeta_2$ are related by  $\zeta_2 = \zeta_1/r = \zeta/r$, where $r$ denotes the ratio between the two frequencies.

The results and analysis presented in the following sections were obtained for a NEMS resonator we have already considered in previous works~\cite{gusso3,gusso19b}. It has length  $l= 5 \, \mu$m, width $b = 800$ nm, thickness $h= 50$ nm, and gaps with $d = 150$ nm. We note that similar results are expected for both larger (MEMS) or smaller devices~\cite{dantas18}. The resonator is assumed to be made of silicon whose Young modulus is $E = 170$ GPa and the density $\rho = 2.3 \times 10^3$ kg m$^{-3}$.  

\section{Generating Pseudo-random Numbers} 
\label{prng}

As observed in Refs.~\cite{dantas18,gusso19c,gusso3} the equation of motion [Eq.~(\ref{odeins})] can present three different kinds of dynamical behavior: (i) a periodic motion with a single or multiple periods; (ii) a pull-in regime, which is an unstable solution corresponding to a situation where the beam collides with the fixed electrodes and (iii) a chaotic regime of oscillation. The regimes (i) and (iii) are identified by the values taken by the maximum Lyapunov exponent $\lambda$. In the first situation we have $\lambda<0$ and for the chaotic regime this exponent is necessarily positive \cite{strogratz}.

Since we are interested in the generation of pseudo-random numbers, it is the chaotic regime that is relevant to us. The random numbers are associated with the displacements of the beam. Such displacements can be experimentally obtained in many different ways, for instance as changes in the capacitance or using strain gauges~\cite{younis11}. The main idea is to search within the space parameter $(V_{AC},V_{DC},\zeta)$ a region in which chaotic dynamics can be obtained. As discussed in~\cite{gusso3}, the proposed system displays robust chaos, i.e. we have a large and continuous domain of points in the space parameter where we can find $\lambda>0$. Following the ideas of Phatak \& Suresh and P-H.~Lee et. al.~\cite{phatak,phlee04}, used to analyze the randomness in logistic maps, we will generate sets of numbers collecting the position of the oscillating beam periodically, with the resonator operating in a chaotic regime obtained for parameters rendering the largests positive values for the Lyapunov exponent.

\begin{figure}[t!]
\begin{center}
\includegraphics[scale=1.0]{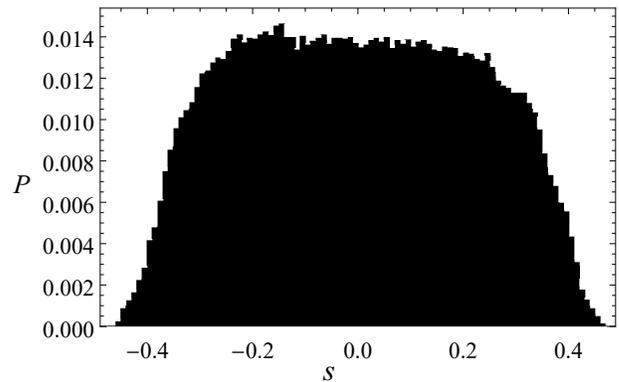}
\caption{Probability distribution $P$ of the beam's positions $\{s\}$ collected at time intervals $T=1.4(2\pi/\zeta)$.}
\label{fig2}
\end{center}
\end{figure}

In what follows we study the generation of random numbers for the resonator operating with $V_{AC}=0.4$~V, $V_{DC}=17.5$~V and $\zeta=0.41$ with the ratio between the frequencies being $r=\sqrt{2}$. Results for other sets of parameters are going to be discussed in Section~\ref{other}. Equation~(2) is numerically solved and we collect the values for the beam position $s$ in times that are multiples of a certain period $T$. We have considered $T=1.4(2\pi/\zeta)$, where the factor $1.4$ was shown to result in more evenly distributed values of $s$. We have observed that, in general, sampling the beam position at periods that are multiple of neither the two excitation frequencies favors the generation of good random numbers. With this procedure we obtained a set of values $\{s\}=\{s_1,s_2,s_3,...\}$ spread over a non-symmetrical domain. The relative distribution of $\{s\}$ obtained in this manner is shown in Fig.~\ref{fig2}. A very distinctive feature of this distribution is that there is a large central region with a quite homogeneous probability. This is in sharp contrast with the distribution of the collected physical variable of other sources of randomness which tend, in the best case, to follow a Gaussian distribution. An ideal source of randomness would have a perfectly homogeneous (or flat) probability distribution of the measured physical parameter. We can thus take advantage of the existence of this more evenly distributed values of $s$ and accept as the initial set of random values the region within $\{s\}$ that presents best homogeneity. We have thus restrained our collection of numbers to values located between $[-0.2,0.2]$. For the final statistical analysis, these numbers are transformed to the more usual interval $[0,1]$ using a linear mapping, thus creating a new set of values $\{x\}=\{x_1,x_2,x_3...\}$.

To obtain our final set of values, which we will investigate if can or cannot qualify as a genuine collection of random numbers, we used a delay in order to eliminate any residue of correlations among our set of values since they were generated from the positions of an equation of motion, and therefore they probably keep some correlation among them. This procedure is the same already employed with impressive results in Ref.~\cite{phatak} for the logistic map. It is performed by taking from the set $\{s\}$ only those values separated by an integer $\tau$, which is the ``delay'', obtaining a sub-set $\{s^1_\tau\}=\{s_1,s_{1+\tau},s_{1+2\tau},...\}$ and performing the linear mapping to the interval $[0,1]$, we get $\{x^1_\tau\}=\{x_1,x_{1+\tau},...\}$. Unlike the authors in Ref.~\cite{phatak}, where they just tested different values for $\tau$, we implement a particular method to get the best choice for this parameter. We consider the $\chi^2$ approach, which is a measure to determine how good a set of values is in order to obtain a flat distribution between $[0,1]$. Certainly, this is a mandatory property for random numbers generated in this interval. Our study concludes that for values above $\tau=12$, we get a distribution sufficiently close to a flat one.

In the next section, we are going to detail the tests performed over this sample of values intending to determine how close to real random numbers they are. It is important to comment that the problem concerning if a set of numbers can or cannot be called random is a question without a solution. Strictly speaking, there is no finite number of tests capable of vouching for a particular set. Instead, those tests are useful to establish if the sample does not qualify as pseudo-random numbers. There is a large number of tests that can be chosen for this purpose. In this work we will focus our attention only on those tests related to statistical properties and physical applications, since one of the most current utilization for random numbers is the numerical simulation of physical problems.

\section{Randomness Tests}
\label{secrand}

To create a set $\{x\}$, where the values $x$ are obtained from a linear transformation made over the original numbers from the set $\{s\}$ and keeping only the values separated by the delay $\tau$ could waste much computational effort if we discard the other values $s$. One way to circumvent this and spare us from longer calculations is composing a set of values $x$ juxtaposing $\tau$ sets of numbers with the elements of each set separated  by the delay $\tau$, i.e. $\{X\}=\{ \{x^1_\tau\}, \{x^2_\tau\}, \{x^3_\tau\}, ...\}$. This set still kept the values apart, at least with a separation $\tau$, and diminishes the correlation between consecutive values. In practical applications of the resonators, this could be easily done using a buffering system.

We performed some evaluations over these sets of numbers to prove two essential characteristics to qualify a sample as random, which are (i) high homogeneity concerning the probability to pick one of those numbers over the interval $[0,1]$ and (ii) low correlation among them, which implies that some sequence of values do not determine the numbers following it. Also, for a HRNG, it is expected that the cycle of random numbers should be as large as possible to avoid the repetition of numbers after a certain period. However, since our values were generated using a chaotic signal, we do not expect this to be an issue because the generated numbers do not have a period.

In order to prove such properties, and other characteristics related to the random numbers, our tests will be divided in two categories: the statistical tests, which are the calculation of the probability that a particular value is located in the domain between 0 and 1, the entropy calculation for the set of numbers considering different number of bins to allocate the values, and the auto-correlation function analysis which will determine how independent those values are from each other. Also, we will test if they satisfy the central limit theorem, a result which is one of the cornerstones for the theory of probabilities, and, finally, we will use the numbers to calculate several orders of statistical moments. The second part of our tests involves the use of the samples generated from the dynamics of the resonator to perform numerical simulations in suitable and well known physical and mathematical problems, in such a way that our results can be directly compared with those already established in the literature and exact ones.

\subsection{Probability Distribution and Entropy}

\begin{figure}[t!]
\begin{center}
\includegraphics[scale=0.3]{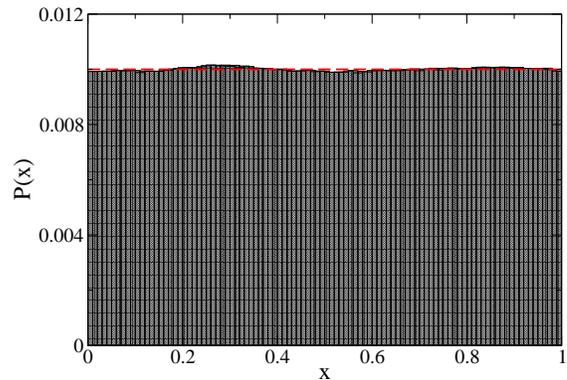}
\caption{Histogram for a set of $N=24 \times 10^6$ numbers generated by the dynamic resonator spread over the interval $[0,1]$ using 100 bins to allocate them.}
\label{fig3}
\end{center}
\end{figure}

Let us take from our set of random numbers $N$ values distributed over the interval $[0,1]$ and group them inside small bins of equal width (of our original interval). We should expect that a genuine set of random numbers will occupy each of those bins with the same probability $P(x)$, and if we have a set large enough, then we will arrive to equally distributed $N$ values among the selected numbers of bins. In our case, we generate a sample $\{X\}$ with $24 \times 10^6$ numbers, and Fig.~\ref{fig3} shows the frequency for those numbers using 100 bins. Strictly speaking, an exact result for random numbers should render us a frequency for all bins equal to 0.01.

\begin{figure}[t!]
\begin{center}
\includegraphics[scale=0.3]{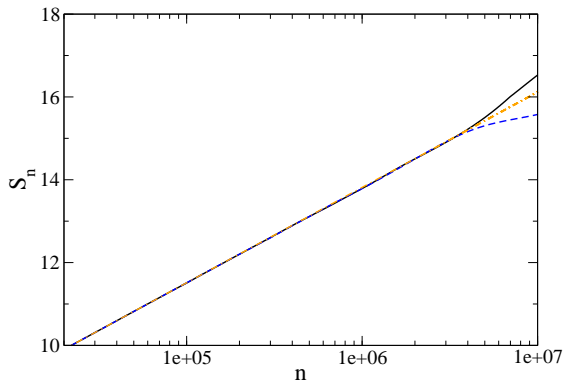}
\caption{Entropy $S_n$ as a function of the number of bins $n$ calculated using numbers obtained by the NEMS dynamic resonator (dashed line) and by the RandomReal routine (solid line). The dashed-dotted line stands from the exact result.}
\label{fig4}
\end{center}
\end{figure}

It is qualitatively perceivable from Fig.~\ref{fig3} that the frequency is almost the same for all values, excluding some minor fluctuations observed due to the finite number of values of our set. To be more precise, we need to quantify this behavior and establish how close our probability distribution is from the flat distribution obtained in an exact scenario. This comparison can be fulfilled noting that in the ideal case, as each bin will group the same number of values, the probability that some value occupies a single bin $i$ will be exactly $p_i=1/n$ where $n$ is the number of bins used to group the set of values spread over some interval. A measure of how equal or not those probabilities are is attained by the concept of entropy as presented in the context of information entropy defined by Shannon~\cite{shannon}, where the entropy $S$ is defined as $S=-\sum_{i=1}^n p_i\ln p_i$. If all probabilities are the same, as should be in a real flat distribution, then $p_i=1/n$ and therefore $S=\ln n$. To check how flat the distribution generated by our sample is, we have separated the interval $[0,1]$ in several different numbers of bins, calculating the probabilities  $p_{i}$ and the entropy $S_n$ for each case. We should expect a linear behavior between $S_n$ and $\ln n$ when the probabilities $p_{i}$ are equal and close enough to $1/n$. The result is displayed in Fig.~\ref{fig4} where we compare the results between our set of numbers with the one obtained using numbers generated by the RandomReal routine from Mathematica~\cite{math}. From Fig.~\ref{fig4}, we can observe that for almost all values of $n$, except for $n \sim 10^7$, the logarithmic behavior of the entropy is found. At the same time, the results obtained using the numbers generated by the RandomReal routine also suffer a deviation from the expected behavior. This fact is not related to a failure in the numbers homogeneity, but instead a simple problem of poor statistic since with $n=10^7$ we have only two numbers per bin.

\subsection{Correlation Tests}

The homogeneity property is not sufficient to vouch for a set of numbers as a genuine random sequence. Actually, we could create samples using some periodic distribution of values and still get a homogeneous probability for all of them. In order to verify that this is not the case, we have also calculated the correlation among those numbers. The ideal case for random numbers is not to display correlation. This implies that averages such as $\langle x_i x_j\rangle$, where $x_i$ and $x_j$ are two numbers from the set $\{X\}$, located at the positions $i$ and $j$ respectively, should obey the relation

\begin{equation}
\langle x_ix_j\rangle = \sum_{i,j} x_ix_j P(x_i,x_j)=\langle x_i\rangle\langle x_j\rangle,
\end{equation}

\noindent since $P(x_i,x_j)=P(x_i)P(x_j)$, because those numbers are independent from each other. 

In order to check how uncorrelated the numbers generated by our mechanism are, we will define the auto-correlation function $C(k)$ as 

\begin{equation}
C(k)=\frac{\langle x_ix_{i+k}\rangle -\langle x_i\rangle\langle x_{i+k}\rangle} {\sqrt{\langle x_i^2\rangle-\langle x_i\rangle^2}\sqrt{\langle x_{i+k}^2\rangle-\langle x_{i+k}\rangle^2}},
\end{equation}

\noindent where $k$ corresponds to a lag which we should compute with different values. If our values are uncorrelated we should have $C(k)\equiv 0$, $\forall k$.

\begin{figure}[t!]
\begin{center}
\includegraphics[scale=0.3]{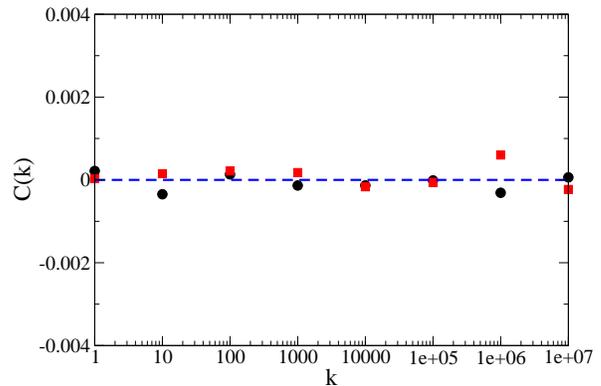}
\caption{Auto-correlation function $C(k)$ as a function of the lag $k$ obtained with values from the NEMS resonator (dots) and the RandomReal routine (squares). The dashed line stands from the exact result.}
\label{fig5}
\end{center}
\end{figure}

It is clear from Fig.~\ref{fig5} that the auto-correlation function is consistent with the random number hypothesis since the auto-correlation function $C(k)$ is very close to zero for a broad domain of lags considered. For comparison, we also show the results obtained using the numbers generated by the RandomReal routine~\cite{math}.

\begin{figure}[t!]
\begin{center}
\includegraphics[scale=0.3]{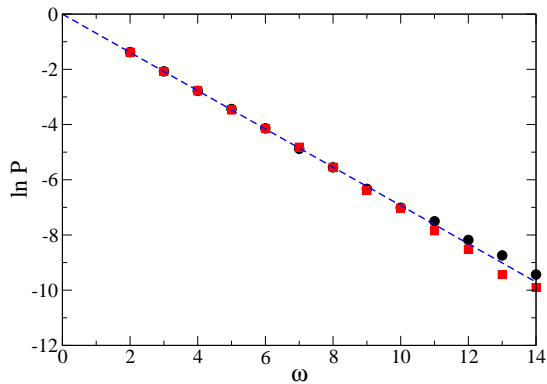}
\caption{Probability to find a sequence of a tuple of length $\omega$ (see text for details). The expected logarithmic behavior is represented by the dashed line. The results for the NEMS resonator and the RandomReal routine are represented by dots and squares, respectively.}
\label{fig6}
\end{center}
\end{figure}

Another test designed to measure correlations is the calculation of probabilities to form tuples~\cite{coins}. Consider that we transform our numbers into zeros or ones following the prescription that $x_i \to 0$, if $x_i \leq 1/2$ and $x_i \to 1$, otherwise. Doing so, we get a set of values like $\{0,0,1,0,1,1,0,...\}$ from which we can calculate, for instance, the probability that two consecutive elements from this set to be $(0,0)$ or $(0,1)$ or any other combination involving a doublet. For a sequence of randomly distributed zeros and ones, we have that the probability to each doublet to appear is $P=2^{-2}$. Actually, we can extend the same logic to a tuple with length $\omega$, $(b_1,b_2,...,b_{\omega})$, with $b_i=0(1)$ and verify that the probability to form any combination of these tuples is $P=2^{-\omega}$. Figure~\ref{fig6} shows these probabilities calculated using our set of random numbers, and they follow the expected behavior of a genuine collection of random zeros and ones. For comparison, we also present the probabilities calculated from the numbers obtained by the RandomReal routine~\cite{math}. Both results coincide with minor discrepancies up to $\omega \approx 14$, probably due to a poor tuples statistics.

\subsection{Central Limit Theorem}

One of the cornerstones of the theory of probabilities is the central limit theorem. This theorem establishes that, under proper circumstances, summing $N$ independent variables $x_i$ generates a new variable $y=\sum_{i=1}^N x_i$ in such a way that when $N \to \infty$ then $y$ becomes a normal distributed value. It means that the resulting variable $y$ will follow a Gaussian (or Normal) distribution given by the expression

\begin{equation}
P(y)=\frac{1}{\sqrt{2\pi N\sigma^2}}\exp\left[-\frac{(y-N\mu)^2}{2N\sigma^2}\right],
\end{equation}

\noindent where $\mu=\langle x\rangle$ is the average and $\sigma^2=\langle x^2\rangle-\langle x\rangle^2$ is the associated variance.

\begin{figure}[t!]
\begin{center}
\includegraphics[scale=0.3]{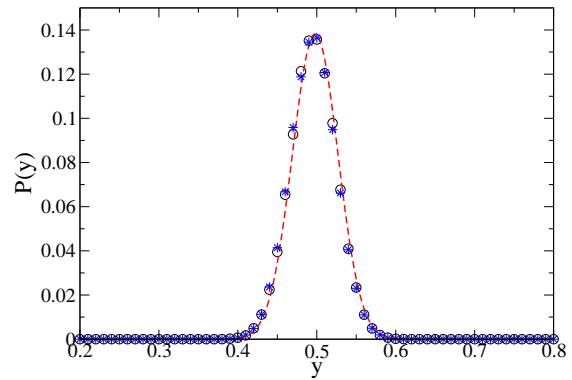}
\caption{Reconstruction of the probability distribution function using $N=100$ numbers to generate each variable $y$ in the set $\{y_i\}$, whose probability has the same form of a Gaussian with average $\mu=1/2$ and $\sigma=1/12$, as should be for numbers located in the interval $[0,1]$ (dashed line). The empty dots represent the result obtained using our set of random numbers and the stars those obtained through the RandomReal routine.}
\label{fig7}
\end{center}
\end{figure}

Figure~\ref{fig7} shows the result for the probability distribution function considering $N=100$ for the generation of each number $y$ in the set $\{y_i\}$ comprised of $24 \times 10^4$ terms. A good coincidence between our values and the exact Gaussian form is observed. Although visual, the apparent coincidence between our result and the exact Gaussian function could be used to argue that our numbers fulfill the central limit theorem, and therefore they could be called truly random. For comparison we show the results obtained with the random numbers generated by the RandomReal routine; also in good agreement.

\subsection{Statistical Moments}

If the result on the previous subsection suggests that our numbers are compatible with the central limit theorem, a more quantitative analysis can be made through the calculation of the moments associated with these values, which are uniquely defined for a Gaussian probability distribution function. The expression for two of the {{\it n}th order} of these moments are given by

\begin{eqnarray}
\mu_n&=&\langle x^n\rangle = \frac{1}{N}\sum_{i=1}^N x_i^n, \nonumber \\
\sigma^2_n&=&\langle x^{2n}\rangle-\langle x^n\rangle^{2} \nonumber \\
&=& \frac{1}{N} \sum_{i=1}^N x_i^{2n} - \left( \frac{1}{N}\sum_{i=1}^N x_i^n \right)^2.
\end{eqnarray}

The above expressions can be simplified if the set $\{x_i\}$ is composed by uniformly distributed numbers limited to the domain $[0,1]$. In such case they result to be exactly

\begin{eqnarray}
\label{eq2}
\mu_n&=&\frac{1}{n+1}, \nonumber \\
\sigma^2_n&=&\frac{n^2}{(2n+1)(n+1)^2}.
\end{eqnarray}

\begin{figure}[t!]
\begin{center}
\includegraphics[scale=0.35]{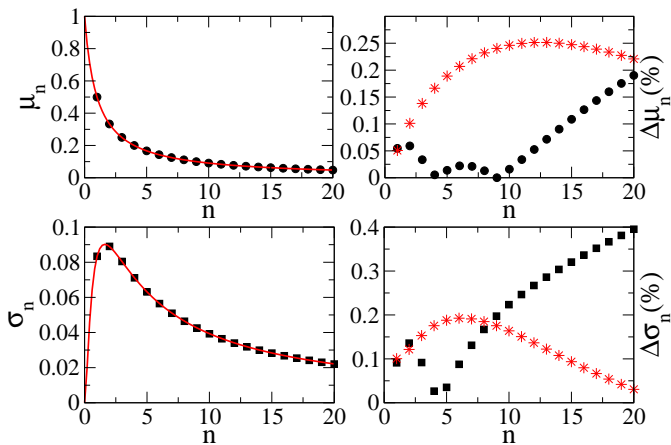}
\caption{Values for the $\it n$th order moments and variances. The left panels compare the exact results from Eq.~(\ref{eq2}) (continuous line) with the ones obtained using the NEMS set (circles and squares). The right panels show the deviation (percentage) of the NEMS and RandomReal routine (stars) results from the exact ones.}
\label{fig8}
\end{center}
\end{figure}

Figure~\ref{fig8} compares the moments obtained with our set $\{X\}$ of $N=24\times 10^6$ terms with the exact results. The left panels show the deviation (percentage) from the exact results for each one of the moments up to order $n=20$. We obtained excellent results for all moments with the deviation never exceeding $0.4\%$. Note that our results are, for some cases, closer to those predicted by Eq.~(\ref{eq2}) than the ones obtained from the set of numbers generated by the RandomReal routine.

\subsection{Random Walk}

Since Ferrenberg et. al.~\cite{ferrenberg92} showed that even well tested pseudo-random generators could fail when used to simulate some physical problems, the statistical tests do not seem enough to assure that some sequences of numbers are authentically random. Although our numbers have passed so far through the statistical tests, we should employ them to simulate physical systems to verify how strong the hypothesis that such numbers are randomly distributed is.

An elementary physical simulation test is the random walk used to study, for instance, diffusion processes. The random walk problem was already vastly investigated~\cite{rudnick04} and have many well-known properties for which we have exact results. We will study random walks in one- and two-dimension regular lattices, with the same step $\ell=1$ and the walker always starting from the origin. After $n$ steps, each step perform with the same probability to any direction, the position of the walker, measured relative to the origin, is $\vec{r}_n$. Two quantities typically associated to a random walk is the average $\langle\vec{r}_n\rangle$, which will vanish, once the walker can move with the same probability to any direction and the so-called mean square displacement given by

\begin{equation}
R^2=\langle|\vec{r}_n|^2\rangle=n\ell^2,
\end{equation}

\noindent which means that the variance of a random walk is proportional to the square root of the number of steps.

\begin{figure}[t!]
\begin{center}
\includegraphics[scale=0.3]{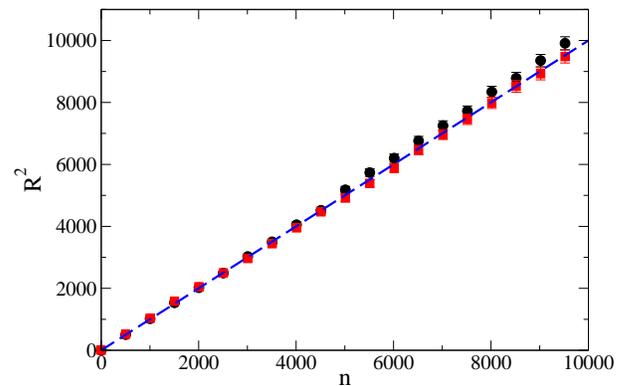}
\caption{Average square distance for the two-dimensional random walk using 2000 walkers of $10^4$ steps all starting from the origin of coordinates. The dots and squares correspond to the results obtained with the NEMS and the RandomReal routine set of numbers respectively.}
\label{fig9}
\end{center}
\end{figure}

Figure~\ref{fig9} shows the mean squared displacement $R^2$ calculated for a two-dimensional walk using our set $\{X\}$ of numbers and the one obtained from the RandomReal routine. We simulate 2000 walks of $10^4$ steps and compare the mean square displacement, $R^2$, to the number of steps to check if the linear behavior is present in our simulation. It is perceivable that up to $n=5\times 10^3$, we have a complete agreement between the exact result and the ones coming from our set of values and those generated by the RandomReal routine. However, starting from this point, our results have a more significant deviation from the expected values in comparison with the sequence provided by the RandomReal routine. Although this deviation is not big enough to disqualify our set of numbers, it may be a sign of residual correlations in the last numbers of the sequence since $R^2$ tend to become bigger than $n$, which is not the case for the numbers obtained via the RandomReal routine.

\begin{figure}[t!]
\begin{center}
\includegraphics[scale=0.3]{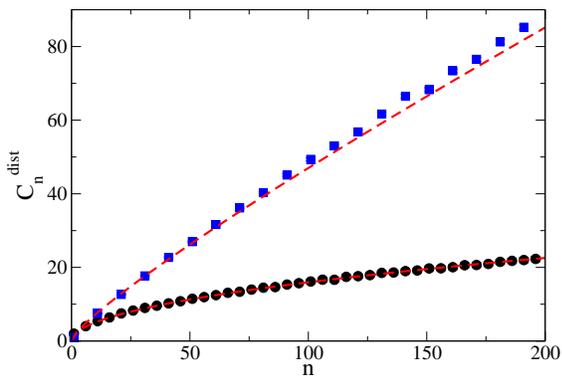}
\caption{Number of distinct visited sites as a function of the number of steps $n$ in the one- (dots) and two-dimensional (squares) random walks. The dashed lines are the expected results.}
\label{fig10}
\end{center}
\end{figure}

Another quantity of interest for the random walk problem is how many different sites are visited by the walker after $n$ steps. For the one-dimensional case, such value is exactly given by $C_n^{\rm 1d}=2\sqrt{2n/\pi}$~\cite{vineyard63} and approximately calculated for the two-dimensional case as $C_n^{\rm 2d}\approx\pi n/\ln(8n)$. Figure~\ref{fig10} compares the results obtained with the NEMS set with the one- and two-dimensional theoretical predictions. The coincidence between numerical and analytical results in the one-dimensional case is quite remarkable. For the two-dimensional case, although minor deviations are observed, especially for large values of $n$, the agreement is excellent.  

On the other hand, a test proposed by Vattulainen~et.~al.~\cite{vatt} to establish if a random generator fails or not is based on the behavior of a proper random walk. If we consider $M$ walkers (all of them starting from the origin of coordinates) and take note of their position after $n$ steps, dividing the space into four quadrants, then a good random generator should be able to produce a final result where the chance that a walk finishes in any quadrant would be simply $E_i=M/4$. Actually, to testify in favor or against the generator we should calculate $\chi^2=\sum_{i=1}^{4}(q_i-E_i)^2/E_i$, where $q_i$ is the fraction of walkers finishing at the ${\it {i}}$th quadrant. Then, to have a random generator with a confidence of $95\%$, we should have $\chi^2<7.815$. The random generator fails if two out of three independent runs fail. Our generator passed this test with $M=1000$ and 500 for the one- and two-dimensional case, respectively, and for several values of steps $n$.

\subsection{Ising Model}

Our last test to establish how useful, or not, the NEMS set can be in order to perform calculations on random processes is going to take place in a simulation of the Ising model~\cite{montroll53,brush67,yeomans92}. We will study first the one-dimensional case because it can be analytically solved. However, we should keep in mind that since our sample $\{X\}$ is limited to $N=24\times 10^6$ numbers, we can only simulate small lattice sizes from which we can obtain acceptable results.

In the one-dimensional Ising model, the only possible magnetization $M$ without an external magnetic field is zero. However, in this case, other quantities are generally used to characterize the thermodynamical state of the system, such as the energy ($E/N_{\rm s}$) and the heat capacity ($C/N_{\rm s}$) per spin, both being functions of the temperature. $N_{\rm s}$ is the number of spins. For the one-dimensional Ising model, these quantities are obtained exactly and given by

\begin{eqnarray}
\label{1dising}
E/N_{\rm s} &=& -\tanh (\beta J), \nonumber \\
C/N_{\rm s} &=& (\beta J) \sech^2 (\beta J),
\end{eqnarray}

\noindent where $\beta=1/k_BT$, with $k_B$ being the Boltzmann constant, and $J$ is a constant with positive value for the ferromagnetic case.

To simulate the one- and two-dimensions Ising models, we considered that each site was occupied by a spin $\sigma_i=\pm 1$ and used the Metropolis algorithm~\cite{binder} in the following way:

\begin{enumerate}

\item We randomly choose a site $i$ using a number $s_i$ from our sample.

\item We switch the signal of the spin $\sigma_i$ and compute the energy of the system with this new configuration using the expression for the Hamiltonian without a magnetic field,

\begin{equation}
\mathcal{H}=-J\sum_{\langle i,j\rangle}\sigma_i\sigma_j,
\end{equation}

\noindent where the sum is over nearest neighbors of the site $i$.

\item If the computed energy is lower than the one obtained with the older configuration, we accept the spin switch. Otherwise, we pick another number $p_i$ from our sample and compare it with the probability $w=\exp(-\beta\Delta E)$, where $\Delta E=E_{\rm new}-E_{\rm old}$. We accept the new configuration only if $w \geq p_i$.

\item At each step we calculate the energy and the magnetization of the system, $M=\sum_i\sigma_i$.

\end{enumerate}

After reaching equilibrium, the quantities of interest $\langle E\rangle, \langle M\rangle$ and $\langle E^2\rangle$ were calculated, with $\langle ...\rangle$ being the average over $n$ repetitions of the Metropolis algorithm. In particular, the heat capacity per spin is related to the quantities obtained from the simulation in the following way,

\begin{equation}
C/N_{\rm s} = \beta^2 [\langle E^2\rangle - \langle E\rangle^2].
\end{equation}

\begin{figure}[t!]
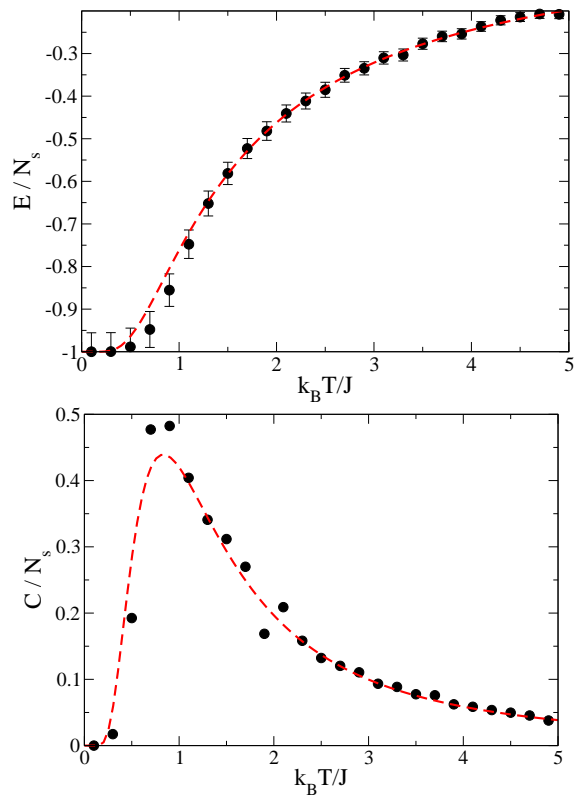

\begin{center}
\includegraphics[scale=0.3]{fig11a.eps}
\includegraphics[scale=0.3]{fig11b.eps}
\caption{Energy (top) and heat capacity (bottom) per spin as a function of the temperature for the one-dimensional Ising model with periodic boundary conditions and considering $N_{\rm s}=100$ spins in the simulation. The dots and the dashed lines correspond respectively to the simulation results using the NEMS set of numbers and to the exact result given by Eq.~(\ref{1dising}).}
\label{fig11}
\end{center}
\end{figure}

In Fig.~\ref{fig11} we show the results obtained for an initially aligned line of spins (with $\sigma_i=1$) with periodic boundary conditions, i.e. a ring of spins. Using our NEMS set of numbers, we were able to compute results up to $N_{\rm s}=100$ spins, using 2000 steps and calculating the average values over $n=300$ repetitions. Although better results can be obtained if we use larger lattices and, consequently, more steps and repetitions, our calculations are limited by the quantity of numbers in our sample. However, as we can see from Fig.~\ref{fig11}, the coincidence between our results and the exact ones expressed by Eq.~(\ref{1dising}) is remarkably good. Except for the low-temperature region where the system takes more time to reach equilibrium since, in the one-dimensional Ising model, the phase transition would occur at $T=0$. We obtained quite similar results with the RandomReal routine set of numbers.

\begin{figure}[t!]
\begin{center}
\includegraphics[scale=0.3]{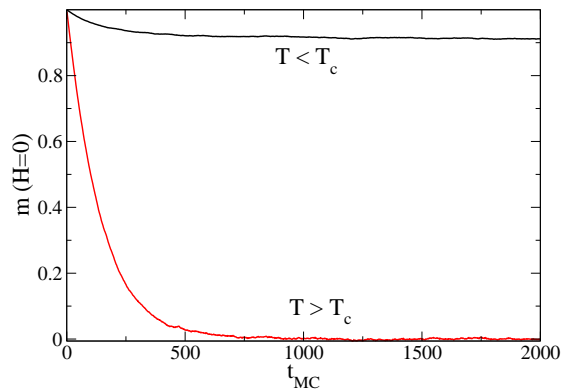}
\caption{Magnetization as a function of the number of Monte Carlo steps for the two-dimensional Ising model, using a $8\times 8$ spin lattice with periodic boundary condition in both directions. The results were obtained for a temperature above and below the critical point.}
\label{fig12}
\end{center}
\end{figure}

We have also performed simulations in a two-dimensional Ising model. We have considered several square lattice sites, with the side going from 2 up to 32. The larger sizes have poorer statistics due to our limited quantity of random numbers in the NEMS sample. Figure~\ref{fig12} shows the time evolution for the magnetization per spin, $m=M/L^2$, in a $8\times 8$ spin lattice with periodic boundary conditions for two temperature values, $T<T_c$ with $m \ne 0$ and $T>T_c$ with $m=0$. The critical temperature $T_c$ for a square lattice without magnetic field was exactly determined by Onsager~\cite{onsager44} as being

\begin{equation}
\frac{k_BT_c}{J}=\frac{2}{\ln(1+\sqrt{2})}\approx 2.26918.
\end{equation}

\begin{figure}[t!]
\begin{center}
\includegraphics[scale=0.3]{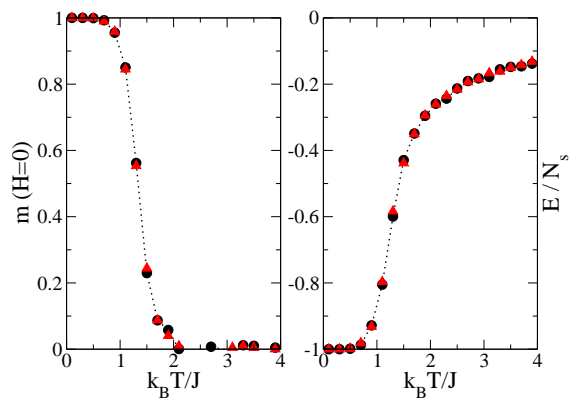}
\caption{Magnetization and energy per spin as a function of the temperature for a $8\times 8$ spin lattice in a two-dimensional Ising model. The circles and the triangles represent the calculations made using the NEMS and the RandomReal set of numbers respectively. The dotted line in the figure is a guide for the eye.}
\label{fig13}
\end{center}
\end{figure}

From the values attained in the steady-state, we have also calculated the energy per spin and the magnetization per spin of the system. Since exact results for a square lattice are only available in the thermodynamic limit, $L \to \infty$, and we were not able to obtain results from lattices large enough to perform a suitable extrapolation to such limit, Fig.~\ref{fig13} only compares our results with those calculated using the RandomReal numbers and as we can see they are quite similar.

\section{Other Samples}
\label{other}

In this section, we will apply the previously discussed tests to other data samples obtained from the NEMS resonator. We considered two cases identified from now on as $S_1=[V_{AC}=0.23V; V_{DC}=17.32; \zeta=0.4]$ and $S_2=[V_{AC}=0.35V; V_{DC}=17.61V; \zeta=0.4]$ for which the resonator dynamics presents a chaotic regime. The samples are both composed by $N=24\times 10^6$ numbers (as in the first set) distributed in the interval $[0,1]$. We will highlight only the main results obtained from those samples.

\begin{figure}[t!]
\begin{center}
\includegraphics[scale=0.3]{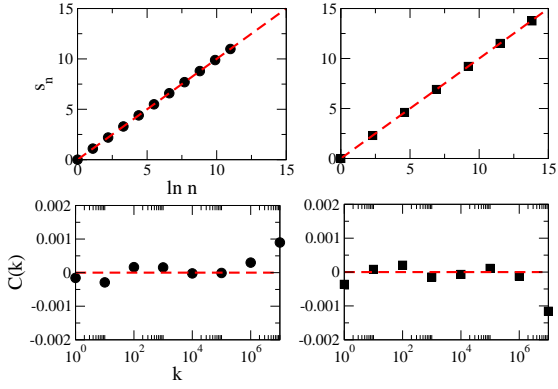}
\caption{Entropy and auto-correlation functions calculated using the samples $S_1$ (left panels) and $S_2$ (right panels). The dashed lines represent the exact analytical result.}
\label{fig14}
\end{center}
\end{figure}

For most of the applied tests both samples have displayed satisfactory results, showing excellent characteristics for homogeneity and also low correlation, as we can see in Fig.~\ref{fig14}, where the entropy and the auto-correlation function are shown. Furthermore, both samples correctly reproduce the results for the number of distinct sites visited in one- and two-dimensional random walks as well as the simulation of the Ising model.

However, there were tests where both samples $S_1$ and $S_2$ revealed a low performance: the calculation of the square distance traveled in the two-dimensional random walk and the test proposed by Vaittulanen~\cite{vatt}. Using the test proposed by Vaittulanen, we observed that only a fraction of our sample could fulfill the $\chi^2$ condition. For the $S_1$ case, only the first $4\times 10^4$ out of $N=24\times 10^6$ numbers used to simulate a two-dimensional random walk with $n=100$ steps and 120 walkers give at least two out of three runs where $\chi^2<7.815$. For the $S_2$ case, this number is bigger, with $6\times 10^5$ values respecting the test condition.

Back to Fig.~\ref{fig14}, we realize that the auto-correlation function has its peak around $k \sim 10^7$, which suggests that the correlation between the first and the last numbers is still high. One possible way to circumvent that problem is the use of an algorithm to shuffle the numbers in each sample. This shuffle mechanism is a technique used to improve the randomness of a series of numbers and can be implemented in several ways. The one chosen in our analysis has the following recipe:

\begin{enumerate}

\item From our sample of numbers $\{x_1,x_2,x_3,...,x_N\}$ we pick-up two of them, $x_i$ and $x_{i+N/2}$, being $i$ an integer between 1 and $N/2$.

\item We define an integer $p= \frac{N}{2} x_{i+N/2}$.

\item We switch the positions of the numbers $x$ located at the positions $i$ and $p$, i.e., $x_i\to x'_p$ and $x_p\to x'_i$.

\item We repeat the above steps $N/2$ times

\item At the end, we have a new list $\{x'_1,x'_2,x'_3...,x'_N\}$.

\end{enumerate}

\begin{figure}[t!]
\begin{center}
\includegraphics[scale=0.3]{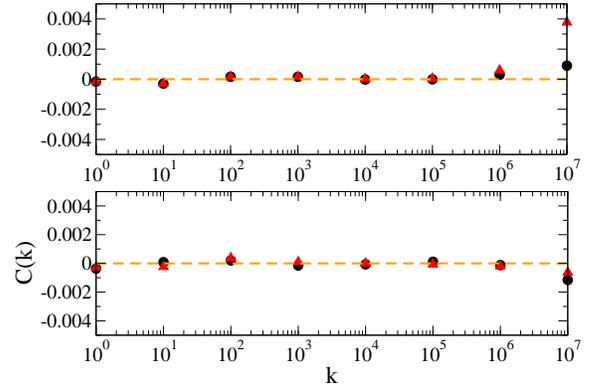}
\caption{Auto-correlation function as a function of the lag size $k$ calculated with the samples $S_1$ (top) and $S_2$ (bottom). The dots and triangles correspond to results obtained with the original and shuffled sample respectively.}
\label{fig15}
\end{center}
\end{figure}

The above procedure was applied to samples $S_1$ and $S_2$ and their shuffled versions ($S'_1$ and $S'_2$) used to calculate the autocorrelation function and to perform the $\chi^2$-test. Figure~\ref{fig15} shows that the shuffle does not affect the sample $S_1$, once the correlation between the first and the last numbers is still high when compared to its unshuffle result. However, the sample $S'_2$ responds better, diminishing that correlation and keeping the other ones close to zero. As a consequence, the $\chi^2$-test is well succeeded up to $2\times 10^6$ numbers, while this number almost does not change for $S'_1$.

\begin{figure}[t!]
\begin{center}
\includegraphics[scale=0.3]{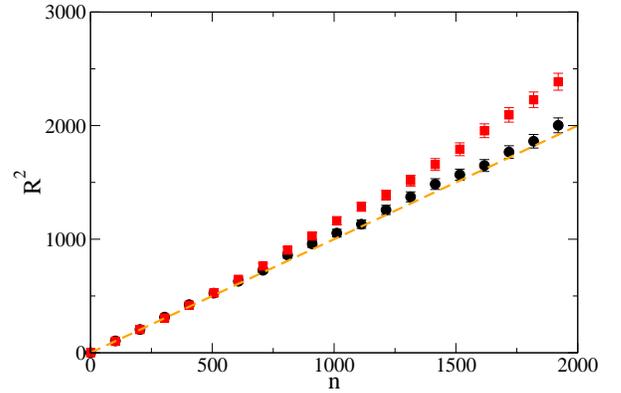}
\caption{Squared average distance performed by a walker in a two-dimensional random walk starting from the origin. The squares and dots were obtained with the $S'_1$ and $S'_2$ samples respectively. The dashed line is the expected result.}
\label{fig16}
\end{center}
\end{figure}

To conclude our discussion, we should comment that the samples $S'_1$ and $S'_2$ were also used to simulate a two-dimensional random walk with $n=2000$ steps and 1000 walkers. We can see from Fig.~\ref{fig16} that $R^2$ shows a bigger deviation from the expected result when the sample $S'_1$ is used, which is the same sample that still keeps a relatively high correlation between the first and the last numbers of the sequence. On the other hand, the $R^2$ results for the $S'_2$ set show a good coincidence with the expected ones, mainly due to the lower correlation between the number's sample. We think that other shuffling strategies can be used to increase the randomness of the samples, consequently improving the results for several cases of chaotic dynamics.

\section{Final Discussions and Conclusion}
\label{conc}

In this work we have shown that taking the beam position as the output signal of a nanoresonator, operating in a chaotic regime, could be used to generate a sequence of values that has properties to certify it as a good random sequence. Therefore, the resonator qualifies as a candidate for a HRNG. The importance of this theoretical result becomes more relevant in view of the recent experimental observation of chaos in a suspended beam MEMS resonator actuated by a single AC voltage~\cite{barcelo19}. The use of two-frequency actuation could, in principle, improve the device reliability as a source of randomness due to the robust chaos it presents.   

The previous statement is based on the results obtained through a series of tests performed over the collection of values obtained from the positions of the resonator beam when it operates in the chaotic regime. Those tests covered statistical properties of the sample as well as numerical simulations in two well-known physical problems, the random walk and the Ising model.

Our main analysis was carried out using a sample generated when the resonator displays chaotic dynamics with a particular choice for the parameters $V_{DC}, V_{AC}$, and the frequency $\zeta$. The set $\{X\}$ used in Section~\ref{secrand} has shown excellent results, passing in all tests performed. The only minor issue has appeared when trying to determine the mean squared displacement in the two-dimensional random walk. For $n\sim 10^4$, small discrepancies from the expected result can be observed, allowing us to speculate that some correlation is still present in the sample. However, despite this minor discrepancy, the system has passed all other tests.

On the other hand, the same thing cannot be said about other sets of values obtained for different points of the parameter space where the chaotic dynamic is present. In Section~\ref{other} we investigated other two samples with inferior results. Nevertheless, a possible solution for this problem could be the use of shuffle protocols to diminish the correlation among the values of each set. This hypothesis was confirmed for one of the analyzed samples, but not for the other one. Perhaps other types of shuffle procedures should have to be tested to improve those samples and obtain the same kind of quality observed for the first set of numbers used in Section~\ref{secrand}.

In this work we have focused the investigation on more fundamental aspects of the NEMS resonator as a source of randomness. A continuous variable, directly related to the beam displacement, was considered as the random variable. In particular, double precision real numbers have been generated and analyzed, but lower precision numbers have also been investigated, leading to the same results. However, for many practical applications we only need to know if the system delivers a sequence of bits that satisfy some criteria. Several strategies to generate the bits sequence can be envisaged. For instance, the generation of multiple bits per sampled position, as would be the case if an analog to digital converter is used in an actual system, or single bits, that can be associated to the beam being at one or the other side of the mean position at the moment of the position sampling. As a sequel to this work we intend to investigate the performance of the binary sequence delivered by the NEMS resonators with robust chaos using tests like DieHard and DieHarder protocols~\cite{marsaglia}, and in applications for cryptography.

\end{document}